\documentstyle[12pt]{article}

\begin{document}
\begin{titlepage}
\begin{flushright}
DO-TH-93/18\\
August 1993
\end{flushright}

\vspace{20mm}
\begin{center}
{\Large \bf
One-Loop Corrections to the Bubble Nucleation Rate at Finite Temperature}
\vspace{10mm}

{\large  J. Baacke\footnote{e-mail~~  baacke@het.physik.uni-dortmund.de}
and V. G. Kiselev\footnote{Alexander von Humboldt Fellow}
\footnote{On leave of absence from Institute of Physics,
220602 Minsk, Byelorussia}
\footnote{e-mail~~  kiselev@het.physik.uni-dortmund.de}} \\
\vspace{15mm}

{\large Institut f\"ur Physik, Universit\"at Dortmund} \\
{\large D - 44221 Dortmund , Germany}
\vspace{25mm}

\bf{Abstract}
\end{center}
We present an evaluation of the 1-loop prefactor
in the lifetime of a metastable state which decays
at finite temperature by bubble nucleation.
Such a state is considered in
one-component $\varphi^4$-model in three space dimensions.
The calculation serves as a prototype application of
a fast numerical method for evaluating the functional determinants
that appear in semiclassical approximations.
\end{titlepage}

%***************************************************************Introduction

\section{Introduction}
\par

The decay of metastable states by bubble nucleation appears in a
large variety of physical contexts. It has received the attention
of particle physicists and cosmologists due to its possible r\^{o}le
in the evolution of the universe.
Since bubble formation is a basic mechanism in the
kinetics of a first-order phase transition a precise determination
of its rate is of prime importance.
The semiclassical approach to bubble formation has been developed
by Langer \cite{La1,La2} and Coleman and Callan \cite{Co1,CaCo}.
The leading factor in the transition rate is determined by the
classical euclidean trajectory. Quantum corrections may however
modify the rate in an significant way. Their evaluation for a realistic
model in three space dimensions is an enterprise that can easily
reach the limits of practical computability. It is therefore
very useful to have method that leads to a fast numerical
algorithm.

We will develop here such a method using as a
simple model the four-dimensional $\varphi^4$-theory
at finite temperature $T$ given by the Euclidean action
\begin{equation}                                                 \label{S}
S(\varphi)=\int_0^{1/T} d\tau \int d^3 x \left( {1\over 2}
\left( \partial _\mu \varphi \right) ^2 + U(\varphi ) \right).
\end{equation}
The field potential $U(\varphi )$ is assumed to have two
non-degenerate minima $\varphi_- =0$ and $\varphi_+ >0$ (Fig. 1).

Any state built on the local minimum $\varphi_-$ is metastable.
Its decay rate per unit volume $\gamma =\Gamma /V$ at sufficiently
high temperature is dominated by the energy $E=ST$ of a fluctuation
which looks like a bubble $\phi(x)$ of the $\phi_+$-phase.
This bubble is in unstable equilibrium between collapse and
unbounded expansion. The tree level approximation
determines the order of magnitude of the decay rate as
$\gamma \approx \exp \{ -E(\phi (x))/T \} $.

Fluctuations around the critical bubble contribute a
pre-exponential fac\-tor to the decay rate which
is known to take the form \cite{La1,La2}
\begin{equation}                                               \label{rate}
\gamma =
\frac{\omega_-  } {2\pi  } \left( \frac{E(\phi)}{2\pi T} \right)^{3/2}
\left| {\cal D} \right| ^{-1/2} \exp \left\{-{E(\phi)\over T}
-S_{ct}\right\}
\end{equation}
to 1-loop accuracy. The coefficient ${\cal D}$ here reads
\begin{equation}                                                \label{det}
{\cal D}(T)
=\frac{det'( -(\partial /\partial \tau )^2 -\Delta + U''(\phi))  }
{det(-(\partial /\partial \tau )^2 -\Delta + U''(0))  }
\end{equation}
The prime in the determinant implies omitting of three
zero modes in it. The temperature dependence of ${\cal D}$ arises from
imposing periodic boundary conditions in the time direction
with a period $1/T$. ${\cal D}(T)$ as introduced in
Eq. (\ref{det}) is ill-defined because of ultraviolet
divergences. As discussed in  \cite{CaCo} they are absorbed
by the counterterm action $S_{ct}$ which has been introduced
in the exponent. It will be specified below.

As we have mentioned above we will present here a fast method
for evaluating the fluctuation determinant (\ref{det}).
The method which we are going to use is based on a well known
theorem \cite{theorem}
that is formulated and proven in an elegant way in Appendix A
of S. Coleman's 1977
Erice Lectures \cite{Co}.
It was also used some time ago in analytical calculations of the determinant
(\ref{det}) in (1+1)-dimensional space in the thin-wall approximation
\cite{KS1,KS2} and in Ref. \cite{S}.
What is new here is application of that idea
to a system in three dimensions, the elaboration of a numerical
method and the discussion of regularization and
renormalization.

We decompose fluctuations around $\phi(x)$
into partial waves, calculate the ratio of determinants $J_l$
of radial operators, using the theorem mentioned above,
and, finally, obtain $\ln {\cal D}$ as $\sum_l (2l+1)\ln J_l$.

In calculating $\ln {\cal D}$ we exclude the divergent perturbative
contributions of first and second order in the external field
created by the critical bubble. The regularized values of
these contributions are then added analytically.
All divergences of $\ln {\cal D}$ appear in the standard
zero-temperature tadpole and fish diagrams.

This paper is organized as follows:
In the next section we specify the form of the potential,
write the equation for the critical bubble
and present our numerical results for $E(\phi)$.
In Sec. \ref{determinant} we describe the calculation of the regularized
fluctuation determinant (\ref{det}).
A possible renormalization scheme is applied to the result
in the Sec. \ref{Renorm}.

%***********************************************************Classical Bubble
\setcounter{equation}{0}
\section{The Tree-Level Energy}
\par

In this section we discuss classical properties of the critical bubble.
The generic one-component $\phi^4$-potential reads
\begin{equation}                                                \label{U}
U(\varphi)=\frac{1  } {2  } m^2 \varphi^2 - \eta \varphi^3
+ \frac{ 1 } { 8 } \lambda \varphi^4
\end{equation}
We choose the same dimensionless variables as in Ref. \cite{DLHLL}, namely
$\vec{x}=\vec{X}/m$, $\tau =u /m$, $\varphi = \frac{m^2  } {2\eta  } \Phi$.
The energy of a time-independent fluctuation then takes the form
\begin{equation}                                                  \label{E}
E(\varphi)= \frac{m^3  } {4\eta^2  }\int d^3 X \left( {1\over 2}
\left( \nabla \Phi \right) ^2 + {1\over2} \Phi^2 - {1\over2} \Phi^3
+{\alpha \over8} \Phi^4 \right).
\end{equation}
where
\begin{equation}                                              \label{alpha}
\alpha = \frac{\lambda m^2  } {2 \eta^2  } .
\end{equation}
The limit $\alpha \rightarrow 1$ corresponds to the thin wall approximation.

The critical bubble is a spherically symmetrical stationary point of
$E$ (\ref{E}) obeying a dimensionless Euler -- Lagrange equation
\begin{equation}                                                \label{eq}
\frac{d^2\Phi  } {dR^2  }+\frac{ 2 } { R } \frac{d\Phi  } {dR  }
-\Phi +{3\over2} \Phi^2 -{1\over2} \Phi^3 =0~~.
\end{equation}
We have solved this equation using the shooting method.
The profiles $\Phi (R)$ are shown in Fig. 2
for some values of the parameter $\alpha$ .

We parametrize the value of $E$ in the same way as in Ref. \cite{DLHLL}:
\begin{equation}                                                 \label{f}
E=\frac{4.851~m^3} {\eta^2  }~f(\alpha)
\end{equation}
The function $f(\alpha)$ is plotted in Fig. 3.

%***********************************************************  Determinant
\setcounter{equation}{0}
\section{Calculation of the Fluctuation Determinant}    \label{determinant}
\par

In this section we discuss a method of computing the ratio of
functional determinants (\ref{det}). which is based on earlier papers
\cite{KS1}-\cite{S}.

The explicit form of the operator in the nominator (\ref{det}) is
\begin{equation}                                                  \label{H}
-\frac{\partial^2  } {\partial \tau^2  }-\Delta+m^2 + V(r)
\end{equation}
with the periodical boundary conditions for the eigenfunctions
$\Psi (\tau, \vec{x}) = \Psi (\tau +1/T, \vec{x})$. The space-dependent
part $V$ (\ref{H}) of $U''(\phi )$ (\ref{det}) reads
\begin{equation}                                                \label{V}
V(r) = U''(\Phi )-m^2
= -6\eta\phi (r)+{3\over2}\lambda \phi^2(r)~~~.
\end{equation}
The free operator in the denominator (\ref{det}) takes the
same form as (\ref{H}), but with $V(r)=0$.

The time-independence and spherical symmetry of the background bubble field
yield a classification of $\omega^2_{n,l,n_r}$,
the eigenvalues of (\ref{H}), with respect to
the number $n$ of their Matsubara frequencies $\nu = 2\pi n T $,
radial quantum number
$n_r$, and angular momentum $l$. One can formally write the ratio of
determinants (\ref{det}) as
\begin{equation}                                              \label{prod}
{\cal D}(T)= \prod_{n=-\infty}^\infty \prod_{l=0}^\infty
\prod_{n_r=0}^\infty \left[
\frac{\omega_{n,l,n_r}}{\omega_{n,l,n_r}^{(0)}} \right]^{2l+1}
\end{equation}
with $\omega_{n,l,n_r}^{(0)}$ standing for the free eigenvalues.

The outline of our calculation is as follows.
We compute first (Sec. \ref{s1}) for each partial wave
the product over $n_r$,
i.e. the ratio of the determinants of the radial operators
\begin{equation}                                                 \label{Dl}
J_l(\nu ) = \frac{det H_{\nu,l}}{det H_{\nu,l}^{(0)} } ;
\end{equation}
here
\begin{equation}                                                  \label{Hl}
H_{\nu,l}=-\frac{d^2  } {dr^2  } - \frac{ 2 } {r  } \frac{d  } {dr  }
+ \nu ^2 + m^2 + V(r)
\end{equation}
and $ H_{\nu,l}^{(0)} $ has the same form,
but without $V(r)$.

As the next step we calculate the product over $l$:
\begin{equation}                                               \label{defF}
F(\nu ) = \sum_{l=0}^\infty (2l+1) \ln J_l(\nu )
\end{equation}
(section \ref{s2})
The function $F(\nu )$ is the sum of all three-dimensional
one-loop one-particle-irreducible diagrams.

In terms of this function the final result reads
\begin{equation}                                               \label{Dfin}
\ln {\cal D}(T) = \sum_{n=-\infty}^\infty F(2\pi n T)~~~.
\end{equation}
This expression is formal because of the ultraviolet
divergences in it. We first evaluate $\ln {\cal D}$ without the divergent
part which is then accounted for in Sec. \ref{s3}.

%******************************************************************
\subsection{Determinants of the Radial Operators}                \label{s1}
\par

In order to find $J_l(\nu )$ (\ref{Dl}) we make use of a known theorem
\cite{theorem,Co} whose statement is
\begin{equation}                                                  \label{th}
\frac{det H_{\nu,l}}{det H_{\nu,l}^{(0)} } =
\lim_{r\to\infty} \frac{\psi_{\nu,l}(r)}{\psi_{\nu,l}^{(0)}(r)}~~~.
\end{equation}
Here $\psi_{\nu,l}$ and $\psi_{\nu,l}^{(0)}$ are solutions to equations
\begin{equation}                                                 \label{eth}
H_{\nu,l}\psi_{\nu,l}=0~~~,~~~~~~H_{\nu,l}^{(0)}\psi_{\nu,l}^{(0)}=0
\end{equation}
and have same regular behavior at $r=0$. More exactly,
the boundary conditions at $r=0$ must be chosen in such a way that
the right-hand side of Eq.(\ref{th}) tends to 1 at $\nu\rightarrow\infty$.

It is convenient \cite{JB} to introduce a function $h(r)$ such as
\begin{equation}                                                  \label{h}
\psi_{\nu,l}=(1+h_l(r))i_l(\kappa r)~~~~,~~~~~~h(0)=0
\end{equation}
when $\psi_{\nu,l}^{(0)}$ is chosen here to be a spherical Bessel function
\begin{equation}
i_l(\kappa r) = \left({2\pi\over \kappa r}\right)^{1/2}
I_{l+{1\over 2}}(\kappa r)~,~~~
\kappa(\nu )^2 =\nu^2+m^2~~~.
\end{equation}
Therefore, by the theorem (\ref{th}), the ratio of determinants
(\ref{Dl}) can be expressed as
\begin{equation}
J_l(\nu ) = (1+h_l(\infty))~~~.
\end{equation}
In terms of the $h$ function the first equation (\ref{eth})
reads
\begin{equation}                                                 \label{eh}
\left(-{d^2\over dr^2}+2\left({i_l'(\kappa r)\over i_l(\kappa r)}m+
{1\over r} \right){d\over dr}\right)h_l(r)=V(r)(1+h_l(r))~~~~.
\end{equation}

It is worth to consider the structure of a perturbation expansion
\begin{equation}                                                  \label{hk}
h_l(r)=\sum_{k=1}^\infty h^{(k)}_l(r)
\end{equation}
in powers of the potential $V(r)$. This entails an analogous
expansion for the ratios $J_l(\nu )$ in the sense that
$J_l^{(k)}=h_l^{(k)}(\infty)$.
The $k$-order contribution $h_l^{(k)}$ obeys an equation
\begin{equation}                                                 \label{ehk}
\left(-{d^2\over dr^2}+2\left({i_l'(\kappa r)\over i_l(\kappa r)}m+
{1\over r} \right) {d\over dr} \right) h_l^{(k)}(r)=V(r)h_l^{(k-1)}(r)~~~,
\end{equation}
$h_l^{(0)}=1$. The same equation is valid when $h_l^{(k)}$ are replaced
by $h_l^{\overline{(k)}}=\sum_{q=k}^\infty h_l^{(q)}$. In this notation
$h_l=h_l^{\overline{(1)}}$.
A Green function that gives the solution to
equation (\ref{ehk}) in the form
\begin{equation}                                               \label{defG}
h_l^{\overline{(k)}}(r)
=-\int_0^\infty dr' r'^2 G_l(r,r')V(r')h_l^{\overline{(k-1)}}(r')
\end{equation}
with the right boundary condition at $r=0$ reads
\begin{equation}                                                   \label{G}
G_l(r,r')= \kappa \left( i_l(\kappa r_<)k_l(\kappa r_>)
\frac{i_l(\kappa r')}{i_l(\kappa r)}
-i_l(\kappa r')k_l(\kappa r') \right)   ~~~~~~.
\end{equation}
Here $r_<=\min \{r,r'\}$, $r_>=\max \{r,r'\}$ and
\begin{equation}
k_l(z)=\left( {2\over \pi z} \right)^{1/2}K_l(z)~~~.
\end{equation}
The first term on the right-hand side of Eq.(\ref{G})
does not contribute to $h_l^{(k)}(\infty)$. The Green function (\ref{G})
gives rise to connected graphs as well as disconnected ones (Fig. 4).
The latter are cancelled in
$\ln (1+h_l(\infty))$ whose expansion in $k$-order connected graphs
$J^{(k)}_{l~con}(\kappa)$ reads
\begin{equation}                                              \label{Jconn}
\ln J_l(\nu ) = \ln (1+h_l(\infty))
=\sum_{k=1}^\infty \frac{(-1)^{k+1}}{k}J^{(k)}_{l~con}(\nu) ~~~.
\end{equation}

This formula is analogous to the expansion of the full functional
determinant in terms of Feynman diagrams
\begin{equation}                                              \label{lnD}
\ln {\cal D}(T)
=\sum_{k=1}^\infty \frac{(-1)^{k+1}}{k} A^{(k)}(T)~~~.
\end{equation}
Here $A^{(k)}(T)$ is the 1-loop  Feynman graph of order $k$
in the external potential $V(|\vec{x}|)$.

Indeed, it is obvious from Eq.(\ref{defG}) that $h_l^{(k)}$ and,
therefore, $J^{(k)}_{l~con}$
are of the order $V^k$. Since the expansion
of $\ln {\cal D}$ in powers of $V$ is unique, we conclude that
\begin{equation}                                              \label{AandJ}
A^{(k)}(T)=\sum_{n=-\infty}^\infty \sum_{l=0}^\infty
(2l+1) J^{(k)}_{l~con}(2\pi nT)~~~.
\end{equation}
One can verify this relation explicitly by expanding the
propagator in $A^{(k)}$ as
\begin{equation}
\int {d^3p \over (2\pi )^3}
\frac{e^{i\vec{p}(\vec{x}-\vec{y})}}{p^2+m^2+\nu^2}=
\kappa \sum_{l=0}^\infty (2l+1)
i_l(\kappa |\vec{x}|) k_l(\kappa |\vec{y}|)
P_l \left(\frac{\vec{x}\cdot \vec{y}}
{|\vec{x}||\vec{y}|} \right)
\end{equation}
and performing the integration
over all angular variables in the $x$-re\-pre\-sen\-ta\-tion.

It is not difficult to check -- making use of a uniform
asymptotic expansion of the modified Bessel functions in (\ref{G}) --
that $J^{(k)}_{con}\sim 1/l^k$ as $l\rightarrow \infty $.
That results in the expected quadratic and logarithmic ultraviolet
divergences in $\ln {\cal D}$
due to the contribution of $h^{(1)}(\infty)$ and $h^{(2)}(\infty)$.
We have computed numerically $\ln {\cal D}^{\overline{(3)}}$
which is the sum
(\ref{lnD}) without first and second order diagrams $A^{(1)}$ and
$A^{(2)}$. It reads explicitly
\begin{equation}                                                \label{sum}
{\cal D}^{\overline{(3)}}(T)
= \sum_{n=-\infty}^\infty F^{\overline{(3)}}(2\pi nT)
= \sum_{n=-\infty}^\infty \sum_{l=0}^\infty  (2l+1)
\left( \ln J_l(\nu ) \right) ^{\overline{(3)}}
\end{equation}
where
\begin{equation}                                               \label{J3b}
\left( \ln J_l(\nu ) \right) ^{\overline{(3)}}
= \left(\ln\left( 1+h_l(\infty)\right)
-h_l^{(1)}(\infty)-\left[h_l^{(2)}(\infty)
-{1\over 2} \left(h_l^{(1)}(\infty)\right)^2 \right]\right)~.
\end{equation}
The terms in square brackets here correspond to the fish diagram
$J^{(2)}_{l~con}$ (Fig. 4). Since all contributions to
$\ln {\cal D}^{\overline{(3)}}$ are
ultraviolet finite, we need no regularization in computing them.
The divergent contributions of the first and second order in $V$
will be considered in Sec. \ref{s3}.

We have determined $h_l(r)$ as solutions of Eq.(\ref{eh}),
and $h_l^{\overline{(2)}}(r)$ as that of Eq.(\ref{ehk})
by Nystrom method.
The values of $h_l^{(1)}(\infty)$,  $h_l^{\overline{(2)}}(\infty)$,
$h_l^{\overline{(3)}}(\infty)$ have been evaluated by performing
integration (\ref{defG}).
Only the last term in the Green function (\ref{G}) contributes here since
$r\rightarrow \infty$.
The ratio of  $h_l^{\overline{(2)}}(\infty)$
found via differential equation (\ref{ehk}) to that calculated
as the integral (\ref{defG}) has been used to control the accuracy.
In order to avoid numerical subtraction that might be delicate
we re-write the term (\ref{J3b}) to
be summed up on the right-hand side (\ref{sum}) in the form
\begin{eqnarray}                                            \label{hren3}
\displaystyle
\left( \ln J_l(\nu ) \right) ^{\overline{(3)}}
&=&\left[\ln(1+h_l(\infty))
-h_l(\infty) + {1\over 2} h_l(\infty)^2 \right]  \nonumber   \\
\displaystyle
&&+~ h_l^{\overline{(3)}}(\infty)
-{1\over 2} h_l^{\overline{(2)}}(\infty)
\left(h_l(\infty) + h_l^{(1)}(\infty) \right) ~~.
\end{eqnarray}
Each of the three terms on the r.h.s. is now manifestly of
order $V^3$.
The subtraction done in the square bracket is exact enough when
the logarithm is calculated with double precision.

In the numerical computation $h_l(\infty)$ is to be replaced,
of course, by $h_l(r_{max})$. We have found that $h_l(r)$
becomes constant within relative deviation of $O(10^{-6})$ for
$r\approx (12 \div 18)/m$ and we have chosen $r_{max}$ in this range
of values.

We have neglected till now the existence of the negative mode
$\omega_{0,0,0}^2<0$ and three zero modes $\omega_{0,1,0}^2=0$.
The former results in negative value of $J_0(\nu )=1+h_0(\infty )$
at $\nu=0$. According to Eq.(\ref{rate}) one has to replace
$\omega_{0,0,0}^2$ by $|\omega_{0,0,0}^2|$.
That implies taking the absolute value of $J_0(0)$ in Eq.(\ref{sum}).

The zero modes manifest themselves by the vanishing of $J_1(\nu )$
at $\nu =0$.
In the numerical calculation this zero was found at
$\omega_{0,1,0}^2 = O(10^{-5}m^2)$.

It can be easily seen \cite{KS1,KS2,S} that
exclusion of the zero modes implies replacing of $J_1(0)$ by
its derivative
\begin{equation}                                                 \label{h'}
J_1'(0)={d\over d(\nu^2)}
h_1(\infty) { \mid}_{\nu=0}~~~.
\end{equation}

In Fig. 5 we present some results for the functions $h_l$.
The values of the
first $h_l^{(k)}(\infty)$ are plotted vs. $2l+1$.
For the terms summed in Eq.(\ref{sum}) we have found good
agreement with the expected behavior $1/(2l+1)^4$.

%*******************************************************************
\subsection{Calculation of ${\cal D}^{\overline{(3)}}$}          \label{s2}
\par

Our next step is performing summation over $l$ in Eq.(\ref{sum}).
It has been done by cutting
the sum at some value $l_{max}$
and adding the rest sum from $l_{max}+1$ to $\infty$ of terms fitted with
\begin{equation}                                               \label{asy}
\ln J_l^{\overline{(3)}} \approx
\frac{Const}{(2l+1)^4}+\frac{Const'}{(2l+1)^6} ~~~.
\end{equation}
The summation was stopped when increasing of
$l_{max}$ by unity did not change the result within some
given accuracy $\delta$. This happened, for example, at $l_{max}=12$ at
$\nu=0,~ \alpha=0.5$. The value of $\delta$ was automatically increased
during the calculation if the corresponding accuracy had not been reached.

The convergence becomes worse at higher $\nu $ or $\alpha $.
The reason is that the asymptotic
behavior (\ref{asy}) sets in at $l\gg (\nu^2+m^2)^{1/2} r_{eff}$
where $r_{eff}$ is the characteristic size of the bubble.
It is of order $1/m$ at small values of $\alpha$ and can be
estimated as $(4/(3(1-\alpha )m)+ O(1) )m^{-1}$ near the thin-wall
limit $\alpha\rightarrow 1$. As the maximal value of the angular
momentum that we have used is $l=30$, the computations have been stopped
somewhere at $\nu \sim 10m$.

The resulting $F^{\overline{(3)}}(\nu )$ is shown in Figs. 6-8.
Its magnitude at $\nu=0$ gives the value of the
infinite-temperature determinant ratio
$\ln {\cal D}^{\overline{(3)}}(\infty)$. To illustrate the efficiency of
the method we note that the evaluation of $F^{\overline{(3)}}(\nu )$ for
one value of $\nu$ takes typically $10\div 30$ sec CPU time on
a standard  PC with 486 processor.

The finite temperature is accounted for in $\ln {\cal D}^{\overline{(3)}}$
which is computed accordingly to Eq.(\ref{Dfin}).
As $F^{\overline{(3)}}(\nu ) \sim 1/\nu ^3$ at high $\nu $, the summation
over Matsubara frequencies is elementary.

%****************************************************************
\subsection{Inclusion of ${\cal D}^{(1)}$ and ${\cal D}^{(2)}$}
\label{s3}
\par

We have found the value $\ln {\cal D}^{\overline{(3)}}$ which is the sum
of all 1-loop diagrams of the third order and higher. Now we add to
the result first- and second-order finite-temperature
Feynman graphs $A^{(1)}(T)$ and $A^{(2)}(T)$ (\ref{lnD})
calculated in the standard technique.
It is convenient to represent both of them if the form
\begin{equation}                                                \label{AT0}
A^{(k)}(T)= \left[ A^{(k)}(T) - A^{(k)}(0) \right] + A^{(k)}(0)
\end{equation}
where the values in square brackets are as follows.
\begin{equation}                                                 \label{A1}
A^{(1)}(T)-A^{(1)}(0) =
{1\over T} \int_0^\infty dr r^2 V(r) Q\left( {m\over 2T}\right)
\end{equation}
with
\begin{equation}                                                  \label{Q}
Q(z) = {z^2\over 8 \pi ^2} \int_1^\infty dy \left( 1+y^2 \right)^{1/2}
\left[\coth (zy) -1 \right] ~~~~.
\end{equation}

The second order gives an UV-finite contribution to $F(\nu )$:
\begin{equation}                                                  \label{F2}
F^{(2)}(\nu) = -4 \int {dq\over q}
\arcsin\left( 1+\left(\frac{2\kappa }{q}\right)^2 \right)^{-1/2}
\left( \int_0^\infty drr V(r)\sin qr \right)^2~~~.
\end{equation}

We have taken $A^{(2)}(T)-A^{(2)}(0)$ (\ref{AT0}) into account
by calculating numerically the difference
\begin{equation}                                                  \label{A2}
A^{(2)}(T)-A^{(2)}(0) = \sum_{n=-\infty}^\infty F^{(2)}(2\pi n T) -
{1\over T} \int_{-\infty}^\infty \frac{d\nu }{2\pi }F^{(2)}(\nu)~~~.
\end{equation}
It is sufficient to make an intermediate cutoff regularization
in order to determine uniquely the difference betwen these
two logarithmically divergent quantities.

All UV divergences have been moved now to the last terms in (\ref{AT0})
which are the standard tadpole and fish diagrams at zero temperature.

To sum up, we have calculated the functional determinant (\ref{det}) as
a sum of following contributions.
\begin{eqnarray}                                             \label{res}
\displaystyle
\ln\left( m^6{\cal D}(T) \right) &=&~\left[  A^{(1)}(T) - A^{(1)}(0) \right]
+ \left[ A^{(2)}(T) - A^{(2)}(0) \right]    \nonumber        \\
\displaystyle
&~&+ \sum_{n=-\infty }^\infty F^{\overline{(3)}}(2\pi n T)
+ A^{(1)}(0) + A^{(2)}(0) ~~~.
\end{eqnarray}
Here the first term is given by Eq.(\ref{A1}), the second one is defined
in (\ref{A2}, \ref{F2}), the sum
of $F^{\overline{(3)}}$ has been calculated numerically, and
two last terms are usual zero-temperature Feynman diagrams (\ref{lnD})
in the external potential $V$.

$A^{(1)}(0)$ and $A^{(2)}(0)$  contain now
all the ultraviolet divergences and have to be  regularized.
The cutoff dependence introduced thereby disappears, however, in the
full 1-loop contribution to the effective action
\begin{equation}   \label{S1loop}
S_{1-loop}=  \frac{1}{2} \ln (m^6{\cal D}(T)) + S_{ct}
\end{equation}
that enters the formula for the transition rate (\ref{rate}).
In the model considered here $S_{ct}$ has the form
\begin{equation}                                             \label{ct}
S_{ct}=\int d^4x \left(
\epsilon \varphi + \frac{1  } {2  } \delta m^2 \varphi^2
- \delta \eta \varphi^3 + \frac{ 1 } { 8 } \delta \lambda \varphi^4
\right)
\end{equation}
A possible scheme for fixing the counterterms is deferred to
the next section.

%********************************************************** Renormalization
\setcounter{equation}{0}
\section{A Possible Renormalization}                         \label{Renorm}
\par

While renormalization requires just standard techniques
it is not straightforward here to select a specific renormalization
precription because the scheme depends strongly on the
physical context in which the first-order phase transition is considered.
Though we consider Eq. (\ref{res}) in its general form as our main
result, we would like to discuss now a possible scheme of fixing
the counterterms.

The scheme is chosen in spirit of renormalization suitable
in consideration of
the electroweak cosmological phase transition.  As the latter is tightly
connected with particle physics, it is appropriate to fix the mass and the
vacuum expectation value of the scalar field. This implies the following
conditions on the temperature-dependent effective potential
$U_{eff}(\varphi ;T)$ and the 1-loop Euclidean propagator
in the true vacuum $D_+(p^2)$:
\begin{eqnarray}                                            \label{onshell}
\displaystyle
D_+(-m^2_+) &=& 0   \nonumber   \\
\displaystyle
U'_{eff}(\phi_+ ;0) &=& 0    \nonumber     \\
\displaystyle
U'_{eff}(0 ;0) &=& 0                \nonumber  \\
\displaystyle
U_{eff}(\phi_+ ;0) &=& U(\phi_+)
\end{eqnarray}
where $m^2_+$ stands for the particle mass in the true vacuum.
The last two conditions set the false vacuum to be at $\varphi=0$
and fix the density of energy stored in it.
They are more specific for our toy model.

In addition to Eqs.(\ref{onshell}) one has to solve one more
system of equations
\begin{eqnarray}                                            \label{class}
\displaystyle
\frac{1  } {2  } m^2 \phi^2_+ - \eta \phi^3_+
+ \frac{ 1 } { 8 } \lambda \phi^4_+ &=& U(\phi_+)  \nonumber \\
\displaystyle
m^2 -3\eta \phi_+ + {1\over 2} \lambda \phi^2_+ &=& 0   \nonumber  \\
\displaystyle
m^2 -6\eta \phi_+ + {3\over 2} \lambda \phi^2_+ &=& 0
\end{eqnarray}
in order to express $m^2$, $\eta $ and $\lambda $ in the classical
bubble energy (\ref{E}) ,(\ref{alpha}) in terms of $m^2_+$, $\phi_+$
and $U(\phi_+)$.

The results of the application of this renormalization scheme are
plotted in  Figs. 9-11. We note that the temperature appears
neither in the field potential (\ref{U}) nor in the renormalization
conditions (\ref{onshell}). All finite-temperature corrections
are therefore contained in the fluctuation determinant.
This results in a linear temperature dependence  due to the first term in
(\ref{res}). Thus, our simple 1-loop approximation fails at very
high temperature. It is not valid also at $\alpha \rightarrow 0$ due to
the high difference of the mass scales of states built on the false
and true vacua. This manifests itself by a logarithmically large
contribution appearing in $S_{ct}$ (\ref{ct}).

%*************************************************************************
\section{Discussion and Conclusion}
\par

The model considered here is only semi-realistic. Nevertheless we
would like to add some remarks on our results.
We find that the correction to the bubble
nucleation rate (\ref{rate}) coming from diagrams of the third
order and higher favors the transition
(i.e. $\ln {\cal D}^{\overline{(3)}}<0$).
The sign of the full 1-loop contribution to the effective action
$S_{1-loop}$ depends on the renormalization scheme.
With our choice (\ref{onshell}) it  enhances $\gamma(T)$ at
high temperature due to the tadpole diagram (\ref{A1}).
This contribution becomes too big at $T\gg m$ and then
the naive application of the 1-loop approximation becomes inconsistent.
This is a manifestation of the
known problem of relating the parameters of the
theory at zero and very high temperatures. Another feature of our results
(Fig. 9-11) is a weak logarithmic singularity at $T=T_{min}$,
the temperature at which one more fluctuation mode of the critical
bubble becomes unstable. Numerically it is unimportant.
Moreover, in a finite region of temperature $T_{min}<T<T_{tunn}$
quantum tunneling has to be taken into consideration.

We have developed here a method for calculating functional
determinants and have tested it under realistic conditions.
In particular we have shown that within this formalism the
problems connected with ultraviolet divergences, zero and
unstable modes can be handled easily and without loss of
numerical accuracy. Moreover the algorithm is so
fast that the whole calculation presented here can be performed
on a standard PC with a 486 processor within one hour
for one value of $\alpha$. The extension of the method to more
complex gauge and fermionic systems is straightforward once the
fluctuation equations are known \cite{JBf,JBSJ}.

%*********************************************************** Aknowledgments
\section*{Acknowledgments}

Work of V.G.K. was supported in its initial stage by Byelorussian
Foundation for Fundamental Research, grant No $\Phi$2-23.
\newpage

%*************************************************************** References

\newpage
%**************************************************************** Captions
\section*{Figure Captions}

\begin{description}

\item[Fig. 1] Potential $U(\varphi )$ (\ref{U}).
It is plotted in dimensionless form which enters the integral (\ref{E}).
The curves are labeled with the value of $\alpha $.

\item[Fig. 2] The bubble-profile functions $\Phi (R)$
in units defined after Eq.(\ref{U}) at $\alpha=0.1,~0.5,~0.8,~0.9$~.
The radius of the thin-walled bubble
$m r_{tw}\approx 3/(4(1-\alpha ))$ is marked with a dashed line
for $\alpha=0.9$~.

\item[Fig. 3] The values of $f(\alpha )$ (\ref{f}) (solid line) and
$-10\omega^2_-/m^2$ (dotted line) vs. $\alpha $.

\item[Fig. 4] The structure of first
$h_l^{(k)}(\infty )$.
The solid line represents the last term in
the Green function (\ref{G}). Dots stand for $V(r)$.

\item[Fig. 5] The values of $J_{l~con}^{(k)}$ at $k=1$ (curve 1),
$k=2$ (curve 2) and $J_{l~con}^{\overline{(3)}}$ (curve 3)
against $2l+1$ in double logarithmic scale
at $\alpha =0$ and $\nu =0$.
The straight-line behavior at large $l$ corresponds to the
expected power law $(2l+1)^{-2k+1}$.

\item[Fig. 6] Absolute value of
$F^{\overline{(3)}}(\nu )$ at $\alpha=0.1$ vs.
$2l+1$ in double logarithmic scale. A logarithmic singularity is
seen at $\nu = \omega_-$. The actual value of $F^{\overline{(3)}}(\nu )$
is negative above this point. In the region $\nu < \omega_-$ it
is complex and has no physical meaning.
The dotted line represents expected relative
error in $F^{\overline{(3)}}(\nu )$ which is estimated as
$5\delta$ (see subsection \ref{s2}).

\item[Fig. 7] The same as in Fig. 6 at $\alpha=0.5$~.

\item[Fig. 8] The same as in Fig. 6 at $\alpha=0.8$~.

\item[Fig. 9] The value of $S_{1-loop}$ (\ref{S1loop}) at $\alpha=0.1$
vs. $T/m$:
The curves 1-3 correspond to the first, second and third terms
in r.h.s.(\ref{res}). The fourth one represents the sum of last two terms
in (\ref{res}) and $S_{ct}$ (\ref{ct}, \ref{onshell}). This
contribution depends on temperature via the factor $1/T$ only.
Curve 5 displays the full result $S_{1-loop}$. The temperature range
is bounded from the left by $T_{min}=0.2301m$ where the result
has a logarithmic singularity.

\item[Fig. 10] The same as in Fig. 9 at $\alpha=0.5$~;
here $T_{min}=0.148m$.

\item[Fig. 11] The same as in Fig. 9 at $\alpha=0.8$~;
here $T_{min}=0.0598m$.

\end{description}

\end{document}